\documentclass[intlimits,twoside,a4paper]{article}

\usepackage[cp1251]{inputenc}

\usepackage[eqsecnum]{cmpj3}

\issue{2023}{26}{4}{43601}
\doinumber{10.5488/CMP.26.43601}
\title[Structural and electronic properties of semimetallic InBi]%
{Study of the structural and electronic properties of semimetallic InBi: first-principles calculation of compound with peculiarities of the electronic structure}
\author[V. V. Pozhyvatenko]{V. V. Pozhyvatenko\orcid{0000-0002-9256-721X}}
\address{Mykolaiv National Agrarian University, 9 Georgiy Gongadze Str., 54008, Mykolaiv, Ukraine}


\date{Received April 22, 2023}

\begin{document}

\maketitle

\begin{abstract}
The electronic properties as well as the structural characteristics and their pressure dependence of the semi-metallic $B10$-structured compound $\mathrm{InBi}$ were investigated.
It is found that the structural values of $\mathrm{InBi}$ calculated in the first-principles calculations reproduce the experimental values worse than those for other heavy III--V pnictides, which are characterized by cubic $B3$ and $B2$ structures,
as well as for IV--VI compounds $\mathrm{SnO}$ and $\mathrm{PbO}$ having the same $B10$ structure.
The low accuracy of the first-principles calculations is a consequence of the peculiarities of the band structure inherent to 
$\mathrm{InBi}$ and not observed in all the other above-mentioned compounds.
To improve the agreement with the experiment, it is proposed to take into account the distortion of the compensated half-metal condition at the highly symmetric points of the Brillouin zone, where the electronic and hole pockets are located.
%
\printkeywords
%
\end{abstract}

\section{Introduction}

The structural and electronic properties of the $\mathrm{InBi}$ compound differ significantly from those of all other III--V pnictides. From the structural point of view, $\mathrm{InBi}$  crystallizes in a lattice that is not typical of the other III--V compounds. From the viewpoint of the electronic structure of InBi, it has features that are not inherent to other III--V pnictides. Therefore, it is interesting to compare both the structural and electronic properties of InBi with other III--V compounds.

Light III--V pnictides under ambient conditions more often possess crystal structure of a zinc blende~($B3$) and is much more seldom a wurtzite ($B4$) one. 
This emphasizes the semiconducting properties of these compounds, in particular, assumes the covalent type of bonding. Considering the pnictides containing the heaviest atoms of III and V groups, namely indium, thallium, and bismuth in their compounds with lighter atoms, we can see that the $B3$ lattice prevails. 

In the $\mathrm{InX}$ compounds, where $\mathrm{X = N, P, As, Sb}$, the $B3$ crystal structure was experimentally defined, except for $\mathrm{InN}$, where the $B4$ structure was found~\cite{CRC}. 
Theoretical calculations in the study of the band structure demonstrate a change in the band gap, namely, its reduction and, accordingly, a decrease in the semiconductor properties of these compounds with an increasing atomic number of element X.

In the $\mathrm{TlX}$ compounds, where $\mathrm{X = N, P, As, Sb}$, the same tendency is possible. Based on theoretical calculations, it is known that more likely these compounds exhibit semimetallic properties. In particular, in reference \cite{MaBeZaFe} the results are given for all such compounds and in reference \cite{Gul} for $\mathrm{TlAs}$ only. However, in reference \cite{WiSchPo} a result was obtained for 
$\mathrm{TlN}$ as a semiconductor with a very small band gap, and in reference \cite{MeDjSaBeRa} it appears that $\mathrm{TlP}$ has semimetallic and even possibly metallic properties.

For $\mathrm{YBi}$, where $\mathrm{Y = B, Al, Ga}$, we see from the calculations of the band structures \cite{FerZao} that they can all have a semi-metallic character.
However, it is worth considering that the density functional theory~(DFT)~\cite{HK} underestimates the band gaps.
As a result of the above calculations, it is likely that only $\mathrm{GaBi}$ can have a semi-metallic character.
The fact that $\mathrm{GaBi}$ is semimetal is also confirmed by the calculations in references \cite{JaWeZh,BaRuRo}. 

\begin{figure}[htb]
\centerline{\includegraphics[width=9cm,height=9cm]{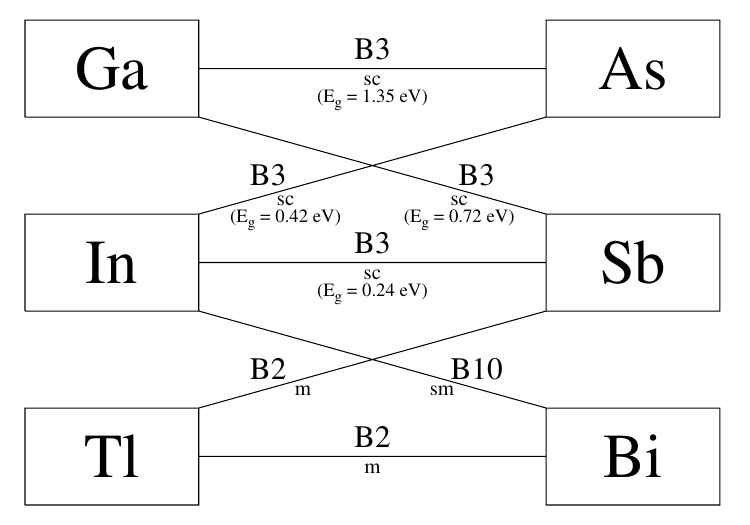}}
\caption{The nature of the chemical bonds of the considered heavy III--V compounds. The line connecting the corresponding elements shows the characteristics of the resulting compound. Here, ``m'' stands for metal, ``sm'' for semimetal, ``sc'' for semiconductor, and $E_g$ for the band gaps of semiconductors \cite{CRC}. $B2$, $B3$ and $B10$ denote the structural type of the compound.}
\label{fig1}
\end{figure}

The heaviest III--V compounds, namely $\mathrm{TlSb}$ and $\mathrm{TlBi}$, are characterized by the $B2$ lattice and metallic bond type. 
The nature of chemical bonding, experimental band gaps in the case of semiconductors, the crystal structure of the heaviest III--V pnictides are shown in figure~\ref{fig1}, where the data are given only for compounds whose atoms are the nearest neighbors in the periodic table. Thus, in this scheme there are no compounds $\mathrm{TlAs}$ and $\mathrm{GaBi}$, experimental data for which are unknown and computational results were discussed above where they were interpreted as semimetals.

\begin{figure}[htb]
\centerline{\includegraphics[width=10cm]{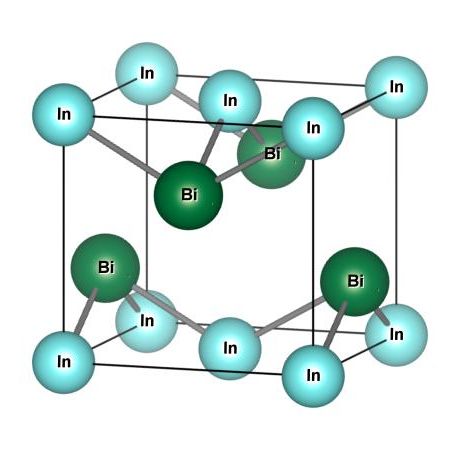}}
\caption{(Colour online) Crystal structure of $\mathrm{InBi}$. Graph produced with VESTA \cite{VESTA}.} 
\label{fig2}
\end{figure}

The properties of $\mathrm{InBi}$ have been studied for quite a long time. The crystal structure was first defined by Binnie \cite{Bin}. This compound crystallizes under ambient conditions in $B10$-lattice in contrast to all other above mentioned pnictides. The $B10$ lattice (see figure~\ref{fig2}) is a tetragonal $\mathrm{PbO}$-type lattice (space group $P4/nmm$). The $\mathrm{In}$ atoms are located in positions 
$(0, 0, 0)$ and $(1/2, 1/2, 0)$ and the $\mathrm{Bi}$ atoms are in positions $(0, 1/2, z)$ and $(1/2, 0, \bar{z})$ with $\bar{z} = 1-z$.  In contrast to the similar $L1_0$ structure, the layer in the plane perpendicular to $c$ forms a corrugated double layer. It should be noted that first-principles calculations reveal in $\mathrm{InBi}$ the $B3$ structure to be thermodynamically advantageous at negligible negative pressures (at lattice stretching), so it makes sense to study this compound in both the $B10$ structure and the $B3$ structure under different thermodynamic conditions.

The known computing works reproduce parameters of crystal lattice of $\mathrm{InBi}$ with insufficient accuracy in first principles approaches. This compound exhibits semimetallic properties, included in the $B3$ structure \cite{JaWeZh}, although  more significantly in the $B10$ structure \cite{FerZao}. Recalling the distortions of the band structures introduced by DFT, we can assume that in the $B3$ structure 
$\mathrm{InBi}$ is still a semiconductor with a narrow band gap, but in the $B10$ structure this compound is a semimetal with a complex Fermi surface which was studied experimentally in references~\cite{Mae,Sch}. The features near the Fermi level~\cite{Mae} undoub\-tedly affect the accuracy of  the first-principles DFT calculations. It is also possible that the insufficient accuracy of calculations is a general trend in their performance for all heavy III--V compounds without exception and does not depend appreciably on the nature of their chemical bonding and, accordingly, on the peculiarities of the electronic structure at the Fermi level.

It turns out that the comparison with other compounds, that possess the same crystal structure as $\mathrm{InBi}$, also reveals significant differences in both structural and electronic properties. It can even be argued that $\mathrm{InBi}$  both structurally and electronically significantly differs from the compounds $\mathrm{SnO}$ and $\mathrm{PbO}$ considered in this article for comparison purposes.
The aforementioned peculiarities of the electronic structure of $\mathrm{InBi}$  in this case also lead to less accuracy of the first-principle calculations for this compound.

In this regard, it is interesting to compare the results of the first-principles calculations of $\mathrm{InBi}$ with those of the IV--VI compounds $\mathrm{SnO}$ and 
$\mathrm{PbO}$.
The known calculations of the structural and electronic characteristics of these compounds \cite{Pelt,Lef,Chen,WPK,Ter} do not reveal the band structure features inherent to $\mathrm{InBi}$.
According to the results of these calculations, $\mathrm{SnO}$ has semimetallic properties over a wide range of temperatures and pressures, while $\mathrm{PbO}$ is a semiconductor.
Note that the compound following $\mathrm{SnO}$, namely $\mathrm{SnS}$, already has semiconductor properties \cite{Ett}.
 
Nevertheless, it is possible to improve the accuracy of the first-principles calculations of $\mathrm{InBi}$  as well as within the framework of the first-principles calculation itself, if we take into account the features of the electronic and hole structure near the Fermi surface, which distort the half-metal compensation.

\section{Computational details}

All calculations in this paper were performed with Quantum ESPRESSO package~\cite{GiBaCa} using the Troullier--Martins pseudopotential \cite{TrMa} in the DFT formalism in the Perdew--Burke--Ernzerhof~\cite{PeBuEr} version of generalized gradient approximation (GGA). The results were fitted to the Birch--Murnaghan equation of state of third order \cite{Mur,Bir}. 
Since $\mathrm{InBi}$ has semimetallic (in both $B3$ and $B10$ phases) properties, the rather complex shape of the Fermi surface should be duly considered in the calculations.
In the case of calculations of the energy characteristics of metals, the fractional occupation number scheme \cite{FuHo} with the smearing parameter $\sigma$ is used, by which states near the Fermi surface are taken into account more accurately and thus the convergence problems of the calculations are solved.
The first-principles calculations show that semimetallic $\mathrm{InBi}$ has a fairly good convergence even without taking into account the smearing ($\sigma=0$), but the contribution of states to the energy of the band structure near the Fermi level is apparently not taken into account accurately enough.
In our calculations we used $\sigma = 0.02$~Ry. Though this is a rather large value, it should be noted that the dependence of the calculation results on the $\sigma$  is quite small, and a markedly large value of this parameter allows the use of smaller Monkhorst--Pack $k$-meshes \cite{MoPa}, which is very convenient when calculating $B10$-structure $\mathrm{InBi}$, since these calculations are computationally expensive due to optimization by three quantities $a$, $c$ (or $\gamma = c/a$) and $z$.  
The Brillouin zone integration was performed using a $8 \times 8 \times 8$ Monkhorst--Pack special $k$-point mesh. The energy cut-off of $70 (280)$ Ry for the wave function (density) was used in all calculations both in this compound  and in $\mathrm{TlSb}$ and 
$\mathrm{TlBi}$. A $20 \times 20 \times 20$ Monkhorst--Pack mesh points were used in the calculations of metallic $\mathrm{TlSb}$ and $\mathrm{TlBi}$. The smearing parameter is the same as in the $\mathrm{InBi}$ calculations.

\section{Results and discussions}

\subsection{Structural properties}

Researches of heavy pnictides in which first of all it is necessary to include the compounds presented in the bottom part of figure~\ref{fig1} are not numerous. The known works are mainly devoted to $\mathrm{InBi}$ which has been computationally investigated for both the $B10$ and the $B3$ structures.   

Table~\ref{tbl1} shows the experimental and the calculated lattice constants for the compounds shown in figure~\ref{fig1}, excluding $\mathrm{InBi}$.
They are obviously divided into the lighter semiconductor III--V compounds with the $B3$ lattice and the heavier metallic pnictides with the $B2$ lattice.  Since a huge number of results are known for the semiconductor compounds mentioned in the upper part of 
figure~\ref{fig1}, the best ones in terms of comparison with the experiment are chosen for this table. This is particularly true for the $\mathrm{GaAs}$ compound for which numerous different calculations have been performed.

\begin{table}[htb]
	\caption{Lattice constants $a$ (in \AA) of heavy III--V pnictides.}
	\label{tbl1}
	\vspace{1ex}
	\begin{center}
		\renewcommand{\arraystretch}{0}
		\begin{tabular}{|l||c|c|c|c|c|c|}
			\hline
			Compound& $\mathrm{GaAs}$&$\mathrm{GaSb}$&$\mathrm{InAs}$&$\mathrm{InSb}$&$\mathrm{TlSb}$&$\mathrm{TlBi}$\strut\\
			\hline 
			Structure& $B3$& $B3$& $B3$& $B3$& $B2$& $B2$\strut\\
			\hline
			\rule{0pt}{2pt}&&&&&&\\
			\hline
			exp.$^{a}$ & 5.65315 & 6.0954 & 6.05838 & 6.47877 & &\strut\\
			\hline
			exp.$^{b}$ & 5.65325 & 6.0959 & 6.0584 & 6.4794 &  &\strut\\
			\hline
			exp.$^{c}$ &  &  &  &  & 3.84 & 3.98\strut\\
			\hline
			GGA-WC $^{d}$ & 5.660 &   &   &    &  &\strut\\
			\hline
			DFT-PAW-HSE $^{e}$ & 5.686 & 6.152 & 6.116 & 6.563 &  &\strut\\
			\hline
			GGA-AM05 $^{f}$ &  & 6.091 &  &  &  &\strut\\
			\hline
			LDA-TB-LMTO $^{c}$ &  &  &  &  & 3.75 & 3.81\strut\\
			\hline
			GGA-PBE $^{g}$ &  &  &  &  & 3.8265 & 3.9117\strut\\
			\hline
			\multicolumn{7}{|l|}{$^{a}$reference \cite{CRC}, 
				$^{b}$reference \cite{AnMaHo}, 
				$^{c}$reference \cite{PaSrSr}, 
				$^{d}$reference \cite{BaRuRo},
				$^{e}$reference \cite{Caro},}\\
			\multicolumn{7}{|l|}{$^{f}$reference \cite{CaSeMe}, 
				$^{g}$ this work.}\\ \hline
		\end{tabular}
		\renewcommand{\arraystretch}{1}
	\end{center}
\end{table}

Table~\ref{tbl2} shows the results for the structural and bulk values of $\mathrm{InBi}$.
The difference between the experimental and the calculated values is noticeable for the lattice parameters 
$a$ and $c$, as well as the tetragonality ratio $\gamma$.
Taking into account that the results for $\mathrm{TlSb}$ and 
$\mathrm{TlBi}$ are significantly improved as a result of the first-principles calculations in this work (0.35\% and 1.72\%, respectively), the result for $\mathrm{InBi}$ remains the worst of all relative errors in the structural parameters of these pnictides.

Let us compare the experimental values of the structural parameters of $\mathrm{InBi}$  with the calculated values. For the calculated values we take the results of the first-principles calculations obtained in this paper. In table~\ref{tbl3}, together with the corresponding calculated values and experimental data from reference~\cite{Bin}, the relative errors for these quantities are given. The largest error is observed for the tetragonality coefficient $\gamma$. We also see a rather large overestimation of the lattice parameter $c$ in the first-principles calculations compared with the experiment. For the nearest-neighbor distances, the distance between the nearest-neighbor In atoms is reproduced the worst. Looking at figure~\ref{fig2}, we notice that the distance depends only on one lattice constant $a$, thus the error accumulates. The other two Bi--Bi and In--Bi nearest-neighbor distances depend on both lattice constants, whose deviations have different signs and are partially compensated. Finally, we see that the unit cell of the corresponding tetragonal structure is excessively elongated along the $z$-axis. It is really bad luck that the experimental tetragonality coefficient is less than unity, while all computational results lead to the values greater than unity, i.e., they destroy the obvious flattening along the $z$-axis.

\begin{table}[htb]
	\caption{Experimental and calculated values of structural and bulk parameters in the $B10$ and $B3$ structures of  $\mathrm{InBi}$.}
	\label{tbl2}
	\vspace{1ex}
	\begin{center}
		\renewcommand{\arraystretch}{0}
		\begin{tabular}{|c|l||c|c|c|c|c|c|c|}
			\hline
			\small Structure&\small Method&\small $a$ (\AA) &\small  $c$ (\AA) &\small  $V_0$ (\AA$^3$) &\small  $\gamma$ &\small   $z$ &\small  $B_0$ (GPa) &\small  $B'_0$\strut\\
			\hline
			\rule{0pt}{2pt}&&&&&&&&\\
			\hline
			\raisebox{-9.2ex}[0pt][0pt]{$B10$}
			&\small exp.$^{a}$ &\small  5.000 &\small  4.773 &\small  119.325 &\small 0.9546 &\small  0.393 &&\strut\\
			\cline{2-9}
			&\small exp.$^{b}$ &\small  5.0118 &\small  4.7790 &\small  120.0396 &\small  0.9536 &\small  0.3924 &&\strut\\
			\cline{2-9}
			&\small exp.$^{c}$ &\small  4.991 &\small  4.776 &\small  118.97 &\small  0.9569 &&&\strut\\
			\cline{2-9}
			&\small GGA$^{d}$ &\small  4.99 &\small  5.03 &\small  125.2475 &\small  1.008 &\small  0.40 &\small  39.95 &\small  4.79\strut\\
			\cline{2-9}
			&\small LDA$^{d}$ &\small  4.81 &\small  4.90 &\small  116.3746 &\small  1.0187 &\small  0.40 &\small  53.70 &\small  4.96\strut\\
			\cline{2-9}
			&\small GGA (relat.)$^{e}$ & \small 5.016 & \small 5.035 &\small  126.6819 &\small  1.0037 &  &\small  42.00 &\strut\\
			\cline{2-9}
			&\small GGA (non-relat.)$^{e}$ &\small  4.974 &\small  5.011 & \small 123.9755 & \small 1.0074 &  & \small 42.00 &\strut\\
			\cline{2-9}
			&\small GGA$^{f}$ & \small 4.888&\small  5.024 &\small  120.0143 & \small 1.028 & \small 0.386 &\small  31.09 & \small 5.51\strut\\
			\hline
			\raisebox{-2.9ex}[0pt][0pt]{$B3$}
			&\small GGA (relat.)$^{e}$ &\small  6.867 &&\small  161.909 &&& \small 30.71 &\strut\\
			\cline{2-9}
			&\small GGA (non-relat.)$^{e}$ &\small  6.901 && \small 164.326 &&& \small 36.84 &\strut\\
			\cline{2-9}
			&\small GGA$^{f}$ & \small 6.772 & & \small 155.263 &&&\small  32.79 &\small  4.89 \strut\\
			\hline
			\multicolumn{9}{|l|}{\small $^{a}$reference \cite{Bin}, 
				$^{b}$reference \cite{JorCla}, 
				$^{c}$reference \cite{Oka}, 
				$^{d}$reference \cite{FerZao}, 
				$^{e}$reference \cite{Zao},
				$^{f}$this work.}\\ \hline
		\end{tabular}
		\renewcommand{\arraystretch}{1}
	\end{center}
\end{table}

The calculated first-principles value for the volume is quite comparable with all experimental values, unlike the lattice constants, because in calculating the volume value, the relative errors of the lattice constants largely compensate each other due to the aforementioned difference in the sign of the cor\-res\-ponding deviations.
It should also be noted that the interlayer parameter $z$ differs markedly from the experimental value.

\begin{table}[htb]
	\caption{Relative errors of the equilibrium structural quantities $B10$-$\mathrm{InBi}$ calculated in this work compared with experiment \cite{Bin}.}
	\label{tbl3}
	\vspace{1ex}
	\begin{center}
		\renewcommand{\arraystretch}{0}
		\begin{tabular}{|c||c|c|c|}
			\hline
			Quantity&exp.&Calc.&Relative error\strut\\
			\hline
			\rule{0pt}{2pt}&&&\\
			\hline
			$a$ (\AA) &5.000&4.888&$-2.24$\strut\\
			\hline
			$c$ (\AA) &4.773&5.024&5.26\strut\\
			\hline
			$\gamma$&0.9546&1.028&7.69\strut\\
			\hline
			$z$&0.393&0.386&$-1.78$\strut\\
			\hline
			$d_{\rm{In-In}}$ (\AA) &3.536&3.456&$-2.24$\strut\\
			\hline
			$d_{\rm{Bi-Bi}}$ (\AA) &3.680&3.641&$-1.06$\strut\\
			\hline
			$d_{\rm{In-Bi}}$ (\AA) &3.126&3.120&$-0.19$\strut\\
			\hline
		\end{tabular}
		\renewcommand{\arraystretch}{1}
	\end{center}
\end{table}

\begin{figure}[htb]
	\centerline{\includegraphics[width=6cm,height=6cm]{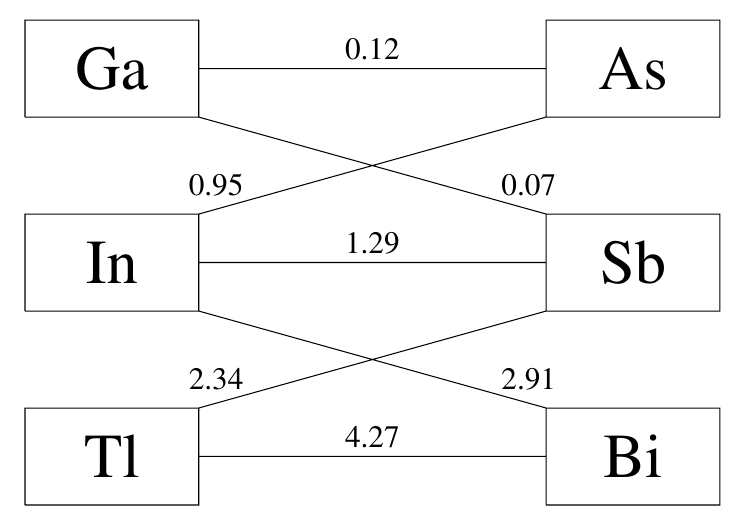}}
	\caption{Relative errors (in the case of $\mathrm{InBi}$, the weighted average relative error) of the lattice constants  obtained in the calculations of other authors in comparison with the experiment 
		(in \%).}
	\label{fig3} 
\end{figure}

The deviations from the experiment for the lattice constants, which are collected in table~\ref{tbl1}, are shown in figure~\ref{fig3}. Thus, these deviations represent the best results obtained in the calculations for these compounds. 
It should be noted that figure~\ref{fig3} shows the value of the weighted average relative error for $\mathrm{InBi}$. 
From figure~\ref{fig3}, it is obvious that the accuracy of the results for heavy pnictides is much worse than the similar accuracy for lighter pnictides. 

\subsection{Pressure dependence of InBi properties}

Let us consider how the structural properties of $\mathrm{InBi}$ change under pressure. 
To compare the results obtained, we use the data from reference \cite{JorCla}, where experimental results of these quantities up to about 2.6~GPa are given.
 
\begin{figure}[htb]
\centerline{\includegraphics{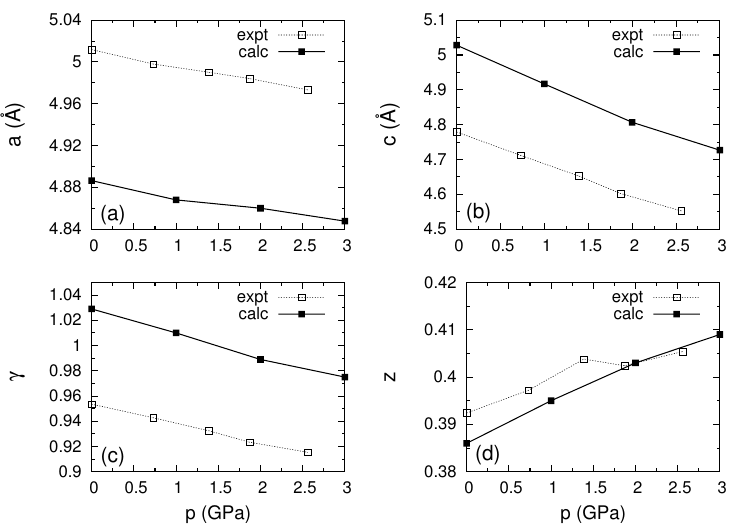}}
\caption{Structural parameters of the $B10$-$\mathrm{InBi}$ crystal lattice versus pressure.}
\label{fig4} 
\end{figure}

Figure~\ref{fig4} shows the change of the structural parameters with increasing pressure. 
It can be seen that the trends with decreasing lattice constants and tetragonality ratio are very similar, given the deviations of the results of the first-principles calculations from the experiment obtained at zero pressure, which practically do not change with increasing pressure.
For the interlayer parameter, the calculated values are close to a more rectilinear dependence with an increasing pressure and show the same increasing trend as the experiment. 

\begin{figure}[htb]
\centerline{\includegraphics{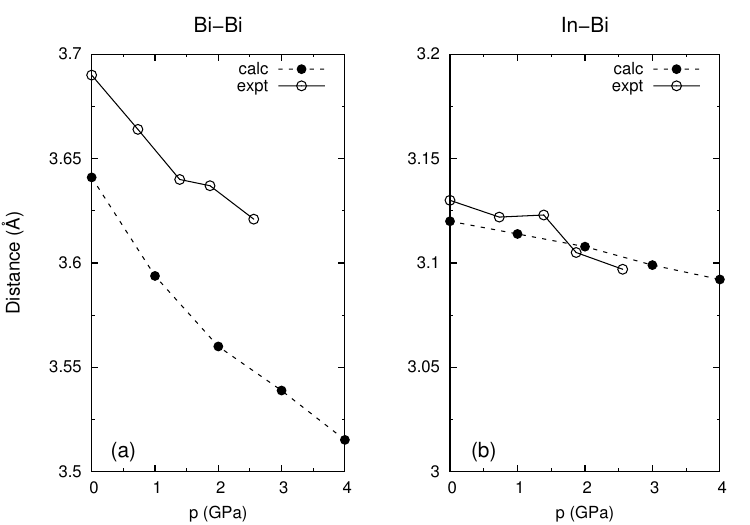}}
\caption{$\mathrm{Bi}$--$\mathrm{Bi}$ and $\mathrm{In}$--$\mathrm{Bi}$ nearest-neighbor distances in the $B10$-$\mathrm{InBi}$ crystal lattices versus pressure.}
\label{fig5} 
\end{figure}

Figure~\ref{fig5} demonstrates the dependence of the distances of the nearest neighbors of bismuth-bismuth and indium-bismuth atoms on pressure. 
Again, a more straightforward dependence of the calculated values compared to the experimental ones is noticeable.
The trends are quite similar, but the bismuth-bismuth distance deviates slightly with increasing pressure from the straight-line trend, thus more reminiscent of the experimental one in which there is a kink in these dependencies at about 1.5 GPa. The dependence for the indium-indium distance, which is the most distorted, coincides with that for the lattice parameter~$a$ to within a multiplier 
(see figure~\ref{fig4}).

\subsection{Electronic properties in heavy III--V compounds with a non-zinc blende structure}

The compounds considered in this section ($\mathrm{InBi}$, $\mathrm{TlSb}$, $\mathrm{TlBi}$) have a non-zinc blende structure, which means that they most likely have a non-covalent type of bonding, unlike semiconductors, which are the lighter representatives of the III--V compounds.

\begin{figure}[htb]
\centerline{\includegraphics[height=6cm]{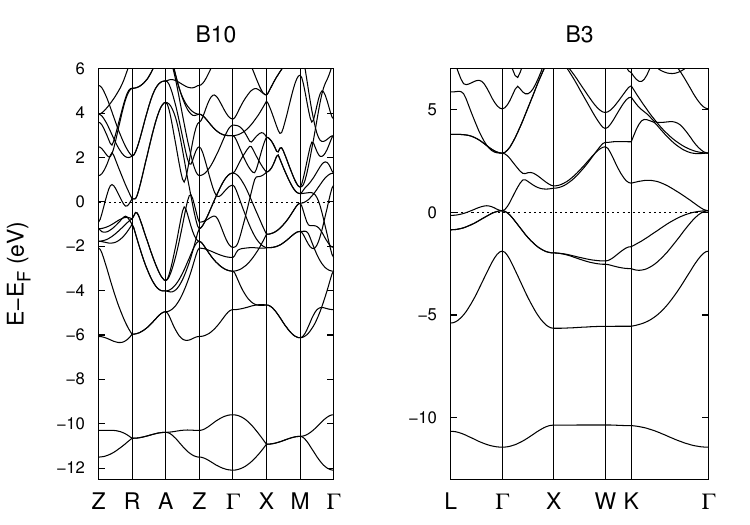}}
\caption{Band structures of $\mathrm{InBi}$ in the $B10$ and $B3$ structures.}
\label{fig6} 
\end{figure}

Figure~\ref{fig6} shows the band structures calculated in this work using first-principles calculations for the $B10$ and $B3$ structures of $\mathrm{InBi}$.
The band structures of $\mathrm{InBi}$ were calculated in earlier works 
\cite{FerZao,Sch,Oka} for $B10$ structure and works \cite{JaWeZh,Sch} for $B3$ structure.

For the $B10$ structure in reference \cite{Sch}, the peculiarities of the band structure near the Fermi level are considered. A visualization of the detected Fermi surface features (electron and hole pockets) is presented in reference \cite{Mae}.
The peculiarities of the band structure obtained in this work are in good agreement with the results given in the above-mentioned papers.
In reference \cite{Oka} the first-principles calculations without and with spin-orbit interactions (SOI) were performed. The results of such consideration lead to the detection of Dirac points.

It is interesting to note that the $B3$ structure is indeed thermodynamically competitive in 
$\mathrm{InBi}$, slightly losing out on the energy to the $B10$ structure under ambient conditions. 
At a small negative pressure $p_{pt} = -0.24$ GPa, corresponding to a stretching of the equilibrium crystal lattice, the $B3$--$B10$ phase transition is computationally detectable. This is apparently a confirmation that it is on $\mathrm{InBi}$ in pnictides that the transition from the $B3$ lattice to other structures and the departure from the predominantly covalent bonding to an increasingly metallic one takes place.

\begin{figure}[htb]
\centerline{\includegraphics[height=6cm]{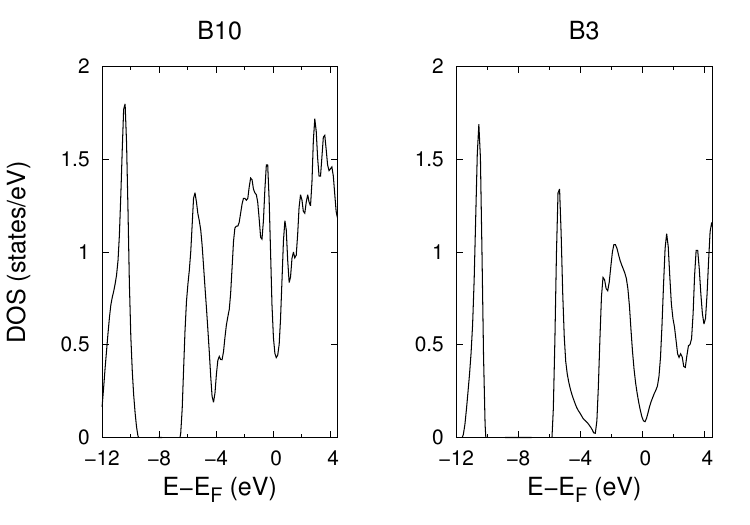}}
\caption{Density of states of $\mathrm{InBi}$ in $B10$ and $B3$ structures.}
\label{fig7} 
\end{figure}

Figure~\ref{fig7} shows the density of states of $\mathrm{InBi}$ for the $B10$ and $B3$ structures. There is a great similarity between these structures in the density of states throughout the entire range presented.  The general appearance of these densities of states is typical of other covalent III--V compounds  \cite{HuCh}. A natural difference is the presence of the bandgap in the III--V semiconductor compounds, which reaches a minimum for compounds with the $B3$ structure in $\mathrm{InSb}$ \cite{HuCh}.

\begin{figure}[htb]
\centerline{\includegraphics[height=6cm]{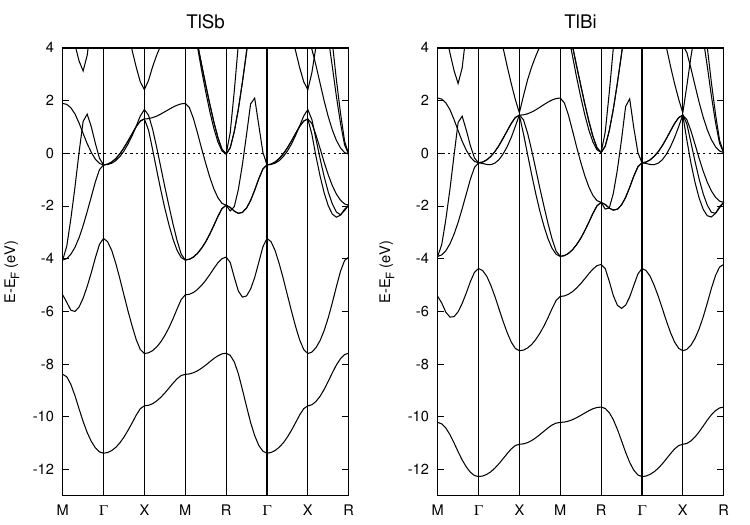}}
\caption{Band structures of $\mathrm{TlSb}$ and $\mathrm{TlBi}$ in the $B2$ structure.}
\label{fig8}
\end{figure}

As for the band structures of $\mathrm{TlSb}$ and $\mathrm{TlBi}$ compounds, which are shown in figure~\ref{fig8}, they quite obviously demonstrate the metallic nature of these compounds.
This confirms that the $\mathrm{InBi}$ compound can be considered as a transition point from semiconductors to metals in the series III--V pnictides with increasing atomic numbers of their constituent elements as shown above in figure~\ref{fig1}.

\subsection{Structural properties of IV--VI compounds with  \textit{B10} structure}

It is interesting to compare the structural characteristics and equilibrium volume of $\mathrm{InBi}$ with the characteristics of the well-known IV--VI compounds $\mathrm{SnO}$ and $\mathrm{PbO}$, where $\mathrm{Sn}$ and $\mathrm{Pb}$ are heavy metals of group IV, which have the structure $B10$, and their position in the periodic table corresponds to the elements of the last two lines in figure~\ref{fig1}.

Table~\ref{tbl4} presents both experimental and computational results of the structural values and volume for the mentioned compounds.
It should be noted that, first, they are very similar to each other, while differing markedly from the values for $\mathrm{InBi}$.
The differences are especially noticeable for the tetragonality parameter $\gamma \approx 1.26$--$1.27$ versus $0.95$ in $\mathrm{InBi}$, as well as for the interlayer parameter $z \approx 0.24$ versus $0.4$ for 
$\mathrm{InBi}$.  

Second, interesting features for $\mathrm{SnO}$ and $\mathrm{PbO}$ compounds relative to $\mathrm{InBi}$, concern a further consideration of their structures.  
Let us introduce general notations for the interatomic distances in the $B10$ structure: $d_1$ for 
$d_{\rm{In-In}}$ and $d_{\rm{O-O}}$; $d_2$ for $d_{\rm{Bi-Bi}}$, $d_{\rm{Sn-Sn}}$ and $d_{\rm{Pb-Pb}}$;  $d_3$ for $d_{\rm{In-Bi}}$, 
$d_{\rm{O-Sn}}$ and $d_{\rm{O-Pb}}$.
The most notable differences in the $B10$ lattices of these structures are as follows.
In $\mathrm{SnO}$ and $\mathrm{PbO}$ $d_2 \approx a$, and in $\mathrm{InBi}$ $d_2 \approx 0.77a$. For 
$\mathrm{SnO}$ and $\mathrm{PbO}$ $d_2/d_3 \approx 1.66$, and for $\mathrm{InBi}$ $d_2/d_3 \approx 1.18$.
Thus, the well-known representation for the $B10$ lattice as alternating $B2$-cells, in which the role of the lattice constant $a$ is played by $d_1$ and $\mathrm{Bi}$ atoms are sometimes above and below the intermediate layer, looks unconvincing in oxide lattices because metal atoms are closer to the oxygen-bearing layers and the corrugated layer splits, creating rather complex basic oxygen layers with metal atoms attached to these layers. 

Third, given the results of \cite{Wang} on the behavior of $\mathrm{SnO}$ under pressure up to 19.3 GPa, with increasing pressure, the characteristics of this compound ($\gamma = 1.1475$, $z=0.2872$ at 19.3~GPa and, respectively, $d_2=0.86a$ and $d_2/d_3 = 1.43$) approach the values for $\mathrm{InBi}$.
However, the lattice changes of the $\mathrm{InBi}$ compound under pressure should be taken into account ($\gamma$ decreases even more, and $z$ tends to 0.5~\cite{Degt}).

\begin{table}[htb]
\caption{Experimental and calculated values of volume and structural parameters in the $B10$ structures of  $\mathrm{PbO}$ and 
$\mathrm{SnO}$.}
\label{tbl4}
\vspace{2ex}
\begin{center}
\renewcommand{\arraystretch}{0}
\begin{tabular}{|c||c|c|c|c|c|c|c|}
\hline
\raisebox{-1.7ex}[0pt][0pt]{\small Quantity}
&\multicolumn{3}{c|}{\small $\mathrm{PbO}$}&\multicolumn{4}{c|}{\small $\mathrm{SnO}$}\strut\\
\cline{2-8}
&\small calc. (LDA$^{a}$) &\small exp.$^{a}$ &\small exp.$^{b}$ &\small calc. (LDA$^{c}$) &\small calc. (LDA$^{d}$) &\small calc. (GGA$^{e}$) &\small exp.$^{f}$\strut\\
\hline
\rule{0pt}{2pt}&&&&&&&\\
\hline
      \small $V_0$ (\AA$^3$)&\small 76.274 &\small 78.543 &\small 79.406 &\small 67.054 &\small 65.811 &\small 75.307 &\small 69.970\strut\\
\hline
     \small  $a$ (\AA)&\small 3.956 &\small 3.965 &\small 3.976 &\small 3.797 &\small 3.76  &\small 3.867 &\small 3.8029\strut\\
\hline
     \small  $c$ (\AA)&\small 4.874 &\small 4.996 &\small 5.023 &\small 4.651 &\small 4.655  &\small 5.036 &\small 4.8382\strut\\
\hline
     \small  $\gamma$&\small 1.232 &\small 1.260 &\small 1.263 &\small 1.225 &\small 1.238  &\small 1.3023 &\small 1.2722\strut\\
\hline
     \small  $z$&\small 0.2403 &\small 0.2368 &\small 0.237 &\small 0.2404 &\small 0.244  &\small 0.234 &\small 0.2383\strut\\
\hline      
\multicolumn{8}{|l|}{\small $^{a}$references \cite{WPK,WP}, 
	$^{b}$reference \cite{Soder}, 
	$^{c}$reference \cite{Pelt}, 
	$^{d}$reference \cite{Meyer}, 
	$^{e}$reference \cite{WW}, 
	$^{f}$reference \cite{PD}. }\\ \hline
\end{tabular}
\renewcommand{\arraystretch}{1}
\end{center}
\end{table}

\begin{figure}[htb]
\centerline{\includegraphics{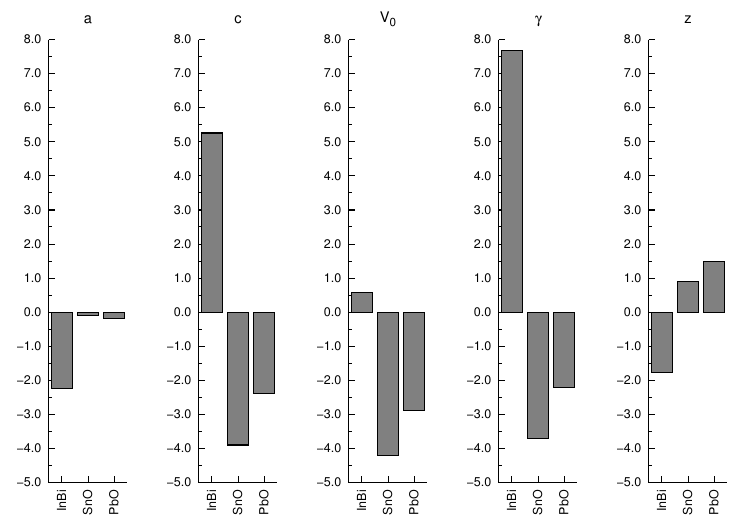}}
\caption{Relative errors (in \%) of $\mathrm{InBi}$, $\mathrm{SnO}$ and $\mathrm{PbO}$ in the $B10$-structure.}
\label{fig9}
\end{figure}

Consideration of the accuracy of the first-principles calculations compared with the experiment will be carried out by comparing the results for $\mathrm{InBi}$ 
(table~\ref{tbl3}) and for IV--VI oxides (table~\ref{tbl4}).
The relative errors of these compounds are visualized in figure~\ref{fig9}, which compares the results from table~\ref{tbl3} for $\mathrm{InBi}$ as well as the results from \cite{Pelt,PD} for 
$\mathrm{SnO}$ and \cite{WPK,WP} for $\mathrm{PbO}$.
Again, we note that the results for the oxides, including the relative errors, are more similar to each other than to the results for $\mathrm{InBi}$.
The largest error in the calculations of IV--VI oxides is obtained for the calculation of the equilibrium volume, because, unlike $\mathrm{InBi}$, the errors for the structural parameters $a$ and $c$ are of the same sign and do not compensate each other.
If we turn directly to the structural quantities, the largest relative error is found when calculating the lattice parameter $c$, then the tetragonality parameter $\gamma$, and then the interlayer parameter $z$. Recall that for $\mathrm{InBi}$ this sequence is $\gamma$, $c$, $a$.

Finally, we can state that the elongation of the unit cell along the $z$-axis affects the properties of the compounds, leading to an increase in the semiconductor properties.
The flattening leads to a semi-metallic state and then, under the applied pressure in 
$\mathrm{InBi}$, it leads to the possible metallization.
It should be noted that the distance between sublayers within a unit cell has a value of $c(1 - 2z)$.
At metallization $z \rightarrow 0.5$ and two sublayers are close to become one layer. Decreasing the $z$ parameter leads to an increase in covalence.

\subsection{Additional consideration of electron states near the Fermi surface}

Figure~\ref{fig9} shows that out of the three $B10$ structures, $\mathrm{InBi}$ has the worst agreement with the experiment, excluding the volume.
It can be seen that the results for $\mathrm{PbO}$ are slightly better than for 
$\mathrm{SnO}$, which can also be explained by the presence of a fairly large band gap in  the former, since according to the results of the calculation of the band structure, 
$\mathrm{PbO}$ is a semiconductor~\cite{WPK,Ter}, and $\mathrm{SnO}$ has semimetallic properties~\cite{Pelt,Lef,Chen}, but does not have the features inherent to $\mathrm{InBi}$ near the Fermi surface, which complicates the calculation of the band structure energy.
The smearing near the Fermi surface used in this paper does not solve the problem of the accuracy of first-principles calculations in $\mathrm{InBi}$.
The choice of a sufficiently large smearing parameter $\sigma = 0.02$ Ry is convenient from the point of view of the speed of convergence of the first-principles calculations, but it is clearly insufficient for complete accounting of the corresponding states, which can be seen from figure~\ref{fig6}, since the value of pockets at points $\Gamma$ and $Z$ exceeds this level of smearing.

The presence of electron and hole pockets (they are also present at the other points in the Brillouin-zone, but are much smaller in size) makes it difficult to accurately calculate the energy of the band structure.
Note that the size of these pockets differs markedly in the calculations made in different papers.
In~\cite{Sch} the electron pocket at the $Z$ point is much larger than the hole pocket at 
the $\Gamma$ point, even taking into account the hole pockets existing in the $\Gamma$--$\Lambda$--$Z$ direction. In~\cite{Oka} it is stated that the sizes of the hole and electron pockets appeared identical, which confirms the concept of compensated semimetal for 
$\mathrm{InBi}$.
However, this only applies to the calculation performed without SOI.
The results of the calculation with SOI from the same paper show the predominance of the pocket size at the $Z$ point as in~\cite{Sch}.
Consideration of this predominance should lead to a curvature of the Fermi surface in the vicinity of the $Z$ point, which should be taken into account when calculating the energy of the band structure.
In this calculation, an attempt is made to take account of the change in the Fermi level at the $Z$ point, based on a comparison of the sizes of the pockets at the $Z$ and $\Gamma$ points, respectively, by adding a contribution that takes into account the change of the Fermi level to a value that provides semimetallic compensation, i.e., the Fermi level shift at the $Z$ point is determined by the difference of the sizes of the pockets at the $Z$ and $\Gamma$ points.
By fitting the dependence $E(\gamma)$ at fixed values of $z$ and finding the energy minimum, we obtain new values of the quantities of interest.

As a result of this calculation, which leads to an additional contribution to the first-principles data, the following values for the structural quantities of $\mathrm{InBi}$ are found:
$a = 4.9529$ \AA $\,(-1\%)$, $c = 4.8464$~\AA $\,(+1.5\%)$, $V_0 = 118.89$~\AA$^3$ 
$\,(-0.4\%)$, $\gamma = 0.9785$ $\,(+2.5\%)$, $z = 0.395$  $\,(+0.5\%)$. 
The relative errors compared to the experiment \cite{Bin} are given in parentheses.
In this approach, the $\gamma$ value is greatly improved.
In addition, the lattice deformation compared to experimental data (where $\gamma<1$) which is inherent to all known first-principles calculations of $\mathrm{InBi}$ is eliminated.

In conclusion, let us make two more remarks.
First, the calculation can be considered not beyond the first-principles approach, because to determine the correction we use the results of the first-principles calculations of the energy levels and band structures calculated in the first-principles approach.
Second, experiments are often performed in ambient conditions and calculations are performed assuming that the absolute temperature tends to zero.
We only know the result of \cite{Takano}, which investigated the temperature dependence of $\mathrm{InBi}$ structural parameters in the range from $-200^{\circ}$C to 
$100^{\circ}$C.
At $-200^{\circ}$C, the following results were obtained in \cite{Takano}:
$a = 4.963$ \AA, $c = 4.848$ \AA {} and $\gamma = 0.977$ and then the relative errors of the results obtained in this work, which additionally take into account the compensation near the Fermi surface, are at most only tenths of a percent.

\section{Conclusions}

The paper considers the structural properties and electronic structure of heavy III--V compound of $\mathrm{InBi}$ using first-principles DFT--GGA--PBE approach. 
The peculiarities of the first-principles calculations in $\mathrm{InBi}$ are much different from those in heavy III--V pnictides with the $B3$ structure, as well as from their related in structure IV--VI compounds $\mathrm{SnO}$ and 
$\mathrm{PbO}$.
For $\mathrm{InBi}$, which is a semimetal with a complex Fermi surface geometry, the accuracy of the first-principles calculations is insufficient to calculate the structural properties with acceptable accuracy.
Heavy III--V pnictides with increasing atomic number show a tendency to move from semiconductor compounds with a cubic $B3$ structure to metallic compounds with a cubic $B2$ structure.
This peculiar transition is also confirmed through the detection in the first-principles calculations at low negative pressures in $\mathrm{InBi}$ of the $B3 - B10$ phase transition, which indicates that the $B3$ phase is still quite competitive in this compound compared to the $B10$ structure.
Interestingly, $\mathrm{InBi}$ exhibits rather semi-metallic properties in both $B3$ and $B10$ structures.
Note also that the $B10$-lattice is related to the B3-lattice due to some remaining 4-coordinated bonds, although structurally it is more similar to the non-cubic version of the $B2$-lattice.

Comparison of the experimental structural data for $\mathrm{InBi}$ with similar data for the known IV--VI compounds ($\mathrm{SnO}$ and $\mathrm{PbO}$), which crystallize in the $B10$ lattice, shows a number of significant differences inherent to the compound $\mathrm{InBi}$.
The most notable difference is that the value of $\gamma$ in IV--VI compounds markedly exceeds 1, and the $z$ value is almost half that of $\mathrm{InBi}$.
That is, the $B10$ lattice for $\mathrm{InBi}$ is much closer to cubic one than in the IV--VI compounds under consideration.
This leads to the established notion that in this case the $B10$ lattice is stretched along the $z$-axis, and the intermediate corrugated layer is actually destroyed, and the corresponding IV-group metal atoms approach the oxygen-containing base layers.
That is, geometrically, these compounds are further away from the representation of the unit cell $B10$-structure through two distorted unit cells of the $B2$ structure, in which the atom is quite close to the center of the unit cell.
As a result, it turns out that the first-principles calculations in the $B10$ lattices of IV--VI compounds describe the structure better, but much worse the equilibrium volume of these compounds.   

Finally, we note that this paper proposes a way to take account of the compensability of the semi-metal in the calculation of the band structure energy and thereby to  improve the agreement with the experiment of the results of this calculation.

\bigskip

\ukrainianpart

\title{Вивчення структурних та електронних властивостей напівметалевого InBi: першопринципний розрахунок сполуки з особливостями електронної структури}
\author{В. В. Поживатенко}
\address{Миколаївський національний аграрний університет, вул. Георгія Гонгадзе 9, 54008, Миколаїв, Україна}

\makeukrtitle

\begin{abstract}
\tolerance=3000%
Досліджено електронні властивості, а також структурні характеристики та їх залежність від тиску напівметалевої сполуки зі структурою $B10$ $\mathrm{InBi}$.
Встановлено, що структурні значення $\mathrm{InBi}$, обчислені в розрахунках з перших принципів, гірше відтворюють експериментальні, ніж для інших важких III--V пніктидів, що характеризуються кубічними $B3$ і $B2$ структурами, а також сполук IV--VI $\mathrm{SnO}$ і 
$\mathrm{PbO}$, що мають ту саму структуру $B10$, що й 
$\mathrm{InBi}$.
Низька точність першопринципних розрахунків є наслідком особливостей зонної структури, властивих $\mathrm{InBi}$ і не спостерігаються у всіх інших вищезгаданих сполуках.
Для покращен\-ня згідності з експериментом пропонується враховувати спотворення умови компенсованого напівметалу у високосиметричних точках зони Бріллюена, де розташовані електронні та діркові кишені.

\end{abstract}

\end{document}